\documentclass[10pt,conference]{IEEEtran}
\usepackage[letterpaper,top=.75in, bottom=1.1in, left=.625in, right=.625in]{geometry}
\usepackage{amsmath,amsfonts}
\usepackage{algorithm}
\usepackage{array}
\usepackage{cite}
\usepackage{verbatim}
\usepackage{stfloats}
\usepackage{tabularx}
\usepackage{lipsum}
\usepackage{url}
\usepackage{graphicx}
\usepackage[caption=false,font=normalsize,labelfont=sf,textfont=sf]{subfig}
\usepackage{multirow}
\usepackage{colortbl}
\usepackage{hhline}
\usepackage{rotating}
\usepackage[hidelinks,colorlinks=true,citecolor=blue]{hyperref}
\usepackage[table,xcdraw,dvipsnames]{xcolor}
\usepackage{tabulary}
\usepackage{amssymb,amsthm}
\usepackage{tikz}
\usetikzlibrary{calc}
\usepackage{tikz-cd}
\usepackage{stackengine}
\usepackage{scalerel}
\usepackage{fp}
\usepackage{color,soul}

\usepackage{enumitem}
\usetikzlibrary{svg.path}
\definecolor{orcidlogocol}{HTML}{A6CE39}
\tikzset{orcidlogo/.pic={
\fill[orcidlogocol] svg{M256,128c0,70.7-57.3,128-128,128C57.3,256,0,198.7,0,128C0,57.3,57.3,0,128,0C198.7,0,256,57.3,256,128z};
\fill[white] svg{M86.3,186.2H70.9V79.1h15.4v48.4V186.2z}
                 svg{M108.9,79.1h41.6c39.6,0,57,28.3,57,53.6c0,27.5-21.5,53.6-56.8,53.6h-41.8V79.1z M124.3,172.4h24.5c34.9,0,42.9-26.5,42.9-39.7c0-21.5-13.7-39.7-43.7-39.7h-23.7V172.4z}
                 svg{M88.7,56.8c0,5.5-4.5,10.1-10.1,10.1c-5.6,0-10.1-4.6-10.1-10.1c0-5.6,4.5-10.1,10.1-10.1C84.2,46.7,88.7,51.3,88.7,56.8z};}}
\newcommand\orcidicon[1]{\href{https://orcid.org/#1}{\mbox{\scalerel*{
\begin{tikzpicture}[yscale=-1,transform shape]
\pic{orcidlogo};
\end{tikzpicture}
}{|}}}}
\theoremstyle{definition}

\usepackage{textcomp}
\usepackage{xcolor}

\usepackage{algpseudocode}

\begin{document}
\title{Lightweight Authenticated Task Offloading in 6G-Cloud Vehicular Twin Networks}
\author{
\IEEEauthorblockN{
Sarah Al-Shareeda\IEEEauthorrefmark{4}\IEEEauthorrefmark{2}\IEEEauthorrefmark{1}, Fusun Ozguner\IEEEauthorrefmark{4}, Keith Redmill\IEEEauthorrefmark{4}, Trung Q. Duong
\IEEEauthorrefmark{5}\IEEEauthorrefmark{6}, and Berk Canberk\IEEEauthorrefmark{3}}

\IEEEauthorblockA{\IEEEauthorrefmark{4}Center for Automotive Research (CAR), Electrical and Computer Engineering Department, The Ohio State University, USA}

\IEEEauthorblockA{\IEEEauthorrefmark{2}BTS Group, Turkey}

\IEEEauthorblockA{\IEEEauthorrefmark{1}Department of AI and Data Engineering, Istanbul Technical University, Turkey}

\IEEEauthorblockA{\IEEEauthorrefmark{5}School of Electronics, Electrical Engineering and Computer Science, Queen's University Belfast, UK}
\IEEEauthorblockA{\IEEEauthorrefmark{6}Faculty of Engineering and Applied Science, Memorial University of Newfoundland, St. John's, NL, Canada}
\IEEEauthorblockA{\IEEEauthorrefmark{3}School of Computing, Engineering and The Built Environment, Edinburgh Napier University, UK}

\{al-shareeda.1, ozguner.2, redmill.1\}@osu.edu, tduong@mun.ca, and b.canberk@napier.ac.uk
}
\markboth{}{}
\maketitle

\begin{abstract}
Task offloading management in 6G vehicular networks is crucial for maintaining network efficiency, particularly as vehicles generate substantial data. Integrating secure communication through authentication introduces additional computational and communication overhead, significantly impacting offloading efficiency and latency. This paper presents a unified framework incorporating lightweight Identity-Based Cryptographic (IBC) authentication into task offloading within cloud-based 6G Vehicular Twin Networks (VTNs). Utilizing Proximal Policy Optimization (PPO) in Deep Reinforcement Learning (DRL), our approach optimizes authenticated offloading decisions to minimize latency and enhance resource allocation. Performance evaluation under varying network sizes, task sizes, and data rates reveals that IBC authentication can reduce offloading efficiency by up to 50\% due to the added overhead. Besides, increasing network size and task size can further reduce offloading efficiency by up to 91.7\%. As a countermeasure, increasing the transmission data rate can improve the offloading performance by as much as 63\%, even in the presence of authentication overhead. The code for the simulations and experiments detailed in this paper is available on GitHub for further reference and reproducibility \cite{sarahalshareeda}.
\end{abstract}

\IEEEpeerreviewmaketitle
\section{Introduction}\label{intro}
Managing computational demands and ensuring secure communication in 6G vehicular networks presents a formidable challenge. As vehicles generate vast amounts of data, offloading computational tasks to cloud or edge servers is essential for maintaining network efficiency. However, this offloading must be meticulously managed to avoid introducing significant latency, particularly in 6G environments that promise ultra-low latency \cite{10440160}. Additionally, secure communication across various network interactions is vital. However, the overhead introduced by authentication frameworks, especially in high-frequency 6G scenarios with limited coverage, can significantly impact the efficiency of offloading \cite{al2020alternating,cheng2023conditional}.

\begin{figure}[!htbp]
\centering
{\includegraphics[width=.75\columnwidth]{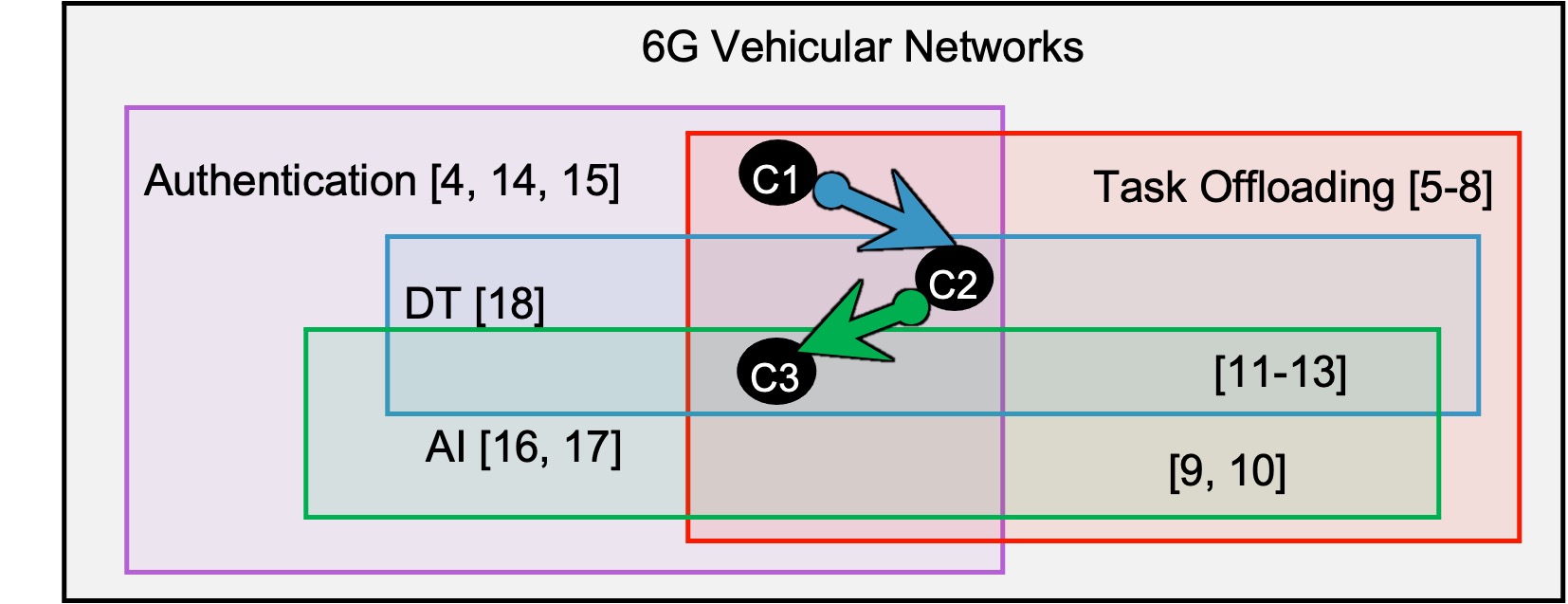}}
\caption{Our Contributions and The Bridged Research Problem.}\label{fig:review}
\end{figure}

Existing research on 6G vehicular networks has primarily addressed task offloading \cite{men2024hierarchical,shen2024slicing,maleki2024handover,liu2024energy,zhou2024edge,fofana2024intelligent,li2023flexedge,tan2024adaptive,hevesli2024task} or authentication \cite{cheng2023conditional,vijayakumar2022anonymous,khowaja2023secure,zhou2021fine,hui2022secure,soleymani2022pacman} in isolation, leaving a critical gap in understanding the combined impact of authenticated task offloading within 6G vehicular networks. The interplay between authentication overhead and offloading efficiency still needs to be explored. This study addresses this gap by developing a unified framework integrating authentication with task offloading within 6G vehicular networks. To facilitate the assessment of the unification, our work leverages two key concepts: Vehicular Twin Networks (VTNs) and Deep Reinforcement Learning (DRL). VTNs, digital replicas of physical networks, utilize Digital Twins (DTs) hosted on cloud or edge servers to mirror vehicles and Roadside Units (RSUs). These DTs provide a comprehensive view of the state of the network, enabling more efficient management of the offloading processes \cite{sarah1, sarah3}. In addition to VTNs, we employ artificial intelligence to model the authenticated offloading latency by formulating it as a DRL task. Specifically, we evaluate the performance of VTNs with IBC-integrated authentication under realistic 6G conditions, focusing on the total latency of the offloading process, from task generation and signing by the vehicle through offloading to the cloud to final reception and verification by the car. We implement a Proximal Policy Optimization (PPO) variant of the DRL model to minimize latency, ensuring efficient task management within the 6G vehicular network. In this context, our main contributions, as illustrated in Fig. \ref{fig:review}, are:

\begin{enumerate}[label=C\arabic*.]
\item Integrating lightweight Identity-Based Cryptographic (IBC) authentication into task-offloading processes within 6G vehicular networks.
\item Defining the model-based VTN architecture where vehicles' twins adapt offloading decisions in the cloud to optimize resource allocation and minimize latency based on real-time network conditions.
\item Developing a PPO-DRL agent that optimizes authenticated task offloading decisions and resource allocation to minimize latency.
\end{enumerate}

The remainder of this article is structured as follows. Section \ref{background} reviews the 6G vehicular networks literature, focusing on authentication and offloading. Section \ref{systemmodel} presents our IBC-based authenticated task offloading architecture for cloud-based VTNs and outlines the offload latency formulation for performance evaluation. Section \ref{hybrid} develops the PPO-DRL model and algorithm. Section \ref{result} covers the simulation settings and discusses the analysis of the results. Section \ref{conc} concludes with key insights and future directions.

\section{Mini Literature Review}\label{background}
Various research has developed robust authentication schemes tailored to the specific requirements of 6G vehicular networks. Cheng et al. \cite{cheng2023conditional} investigate conditional privacy-preserving multi-domain authentication within the Internet of Vehicles in 6G networks. Their framework combines pseudonym management with multi-domain authentication to ensure robust privacy protections without sacrificing performance. Vijayakumar et al. \cite{vijayakumar2022anonymous} propose an anonymous batch authentication and key exchange protocol to reduce the overhead associated with authentication in 6G-enabled vehicular networks' high-mobility scenarios. With the advent of AI, Zhou et al. \cite{zhou2021fine} propose a fine-grained access control approach that utilizes IBC authentication in conjunction with Deep Learning (DL). Their approach effectively manages access control while detecting cyber-attacks with high accuracy. Similarly, Hui et al. \cite{hui2022secure} propose an AI-driven framework for secure and personalized edge computing services in 6G heterogeneous vehicular networks. As the concept of DTs becomes increasingly relevant, Soleymani et al. \cite{soleymani2022pacman} develop a privacy-preserving authentication for cybertwin-based 6G networks. However, they do not incorporate any AI techniques; instead, they focus on the challenges of twin integration.

Aside from authentication, task offloading is another critical focus area in 6G vehicular networks, driven by the need to manage these environments' vast data and computational demands. Hierarchical and multi-layered offloading frameworks are investigated to address the complexities of 6G environments. Men et al. \cite{men2024hierarchical} propose a hierarchical aerial computing framework that utilizes the mobility of drones and edge servers to enhance task offloading and resource allocation in scenarios where ground-based resources are insufficient. Shen et al. \cite{shen2024slicing} focus on resource allocation in Space-Air-Ground integrated vehicular networks through a slicing-based task offloading mechanism. Their framework ensures efficient utilization of heterogeneous resources across different network layers. Adaptation to dynamic network conditions is another essential aspect of task offloading in 6G environments. Maleki et al. \cite{maleki2024handover} introduce a dynamic offloading scheme that is handover-enabled, adjusting to real-time network conditions and vehicle mobility patterns. Liu et al. \cite{liu2024energy} further contribute an energy-efficient joint computation offloading and resource allocation strategy. It balances computational efficiency with energy consumption by offloading tasks to edge and cloud servers.
Utilizing AI, Zhou et al. \cite{zhou2024edge} present an edge offloading that integrates RL with content caching to optimize offloading decisions in the 6G-enabled Internet of vehicles. Similarly, Fofana et al. \cite{fofana2024intelligent} apply DRL for intelligent offloading in vehicular networks. Their method dynamically adjusts the offloading strategies based on real-time computational demands. Integrating DT technology with AI-driven task offloading has been introduced by Li et al. \cite{li2023flexedge}. They introduce FlexEdge, a DT-enabled UAV-aided vehicular edge computing framework. This approach uses DTs to model vehicles and UAVs, optimizing real-time offloading decisions via a PPO-DRL-based algorithm. Similarly, Tan et al. \cite{tan2024adaptive} explore adaptive task scheduling in DT-empowered cloud-native vehicular networks. Hevesli et al. \cite{hevesli2024task} optimize task offloading in DT-assisted edge Air-Ground industrial IoT 6G networks, using DTs and a deep Q-network (DQN) to make real-time decisions on task processing across edge, cloud, or ground resources. 

Existing studies tend to focus on either authentication or task offloading in isolation. Our contribution addresses this gap by developing a unified framework that integrates authenticated task offloading within 6G vehicular networks, models it through vehicular cloud twins, and optimizes its performance using AI techniques, as seen in Fig. \ref{fig:review}.

\section{IBC-based Task Offloading in 6G-VTNs} \label{systemmodel}
\begin{figure}[!htbp]
\centering
{\includegraphics[width=.78\columnwidth]{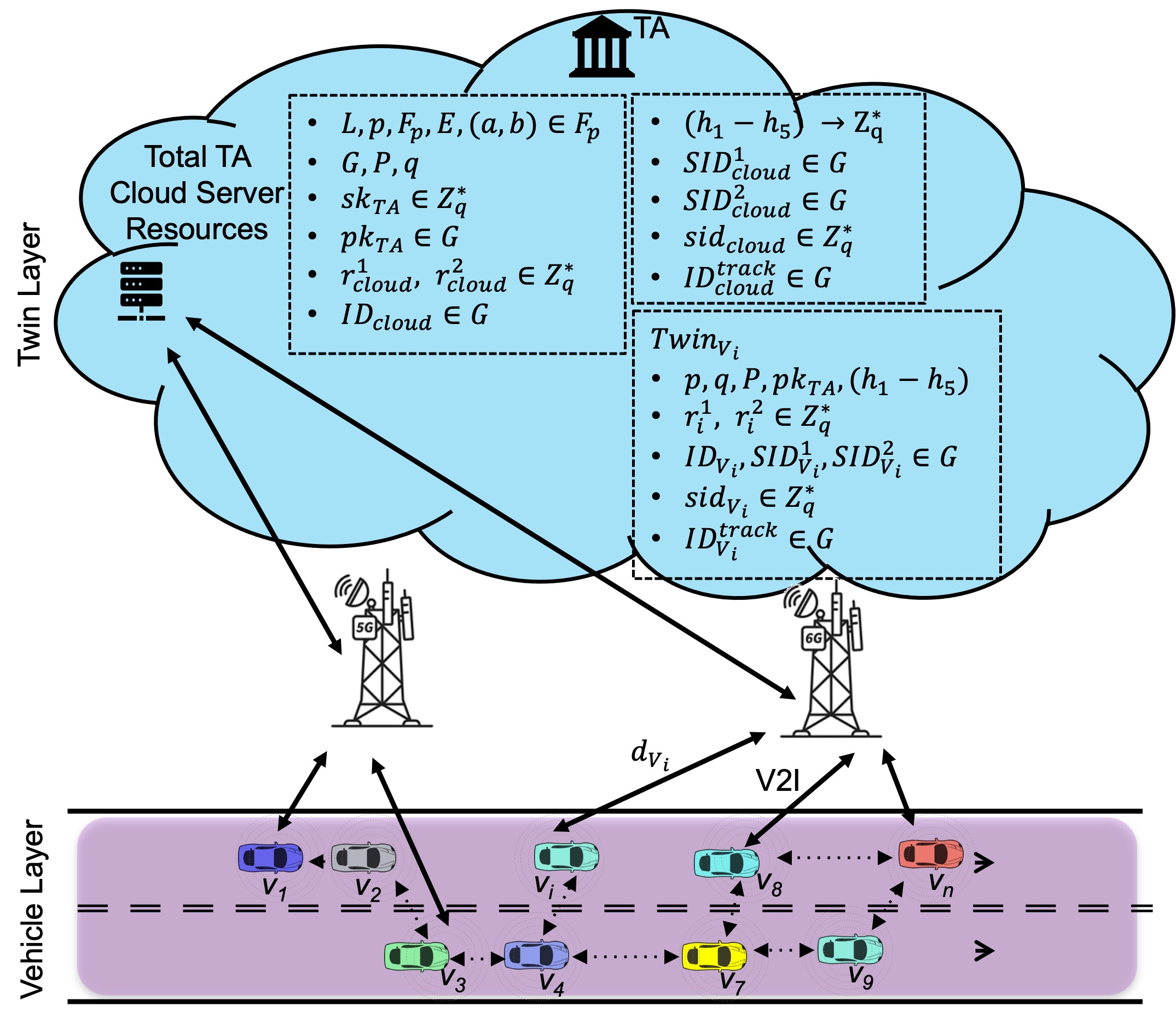}}
\caption{IBC-based Authenticated Cloud-VTN Offloading Architecture.}\label{fig:dep}
\end{figure}

We have developed a cloud-based IBC-authenticated VTN architecture, as shown in Fig. \ref{fig:dep}. The physical layer comprises $n$ vehicles traveling along a highway, each equipped with $f_{V_i}$ GHz of computational resources and moving at $v_{V_i}$ km/h. The architecture's twin layer includes a cloud server with a $F_{cloud}$ GHz computational capacity. Vehicle-to-Infrastructure (V2I) communication between any vehicle $V_i$ and the cloud server occurs at a data rate of $T_{icloud}$ Mbps, reflecting the lower end of the 6G spectrum to account for signal attenuation and the challenges of maintaining stable connections at higher frequencies. The cloud server hosts the virtual counterparts of the vehicles, referred to as $Twin_{V_i}$.

Before deployment, a Trusted Authority (TA) sets the parameters of the IBC authentication \cite{al2018lightweight, al2020alternating} elliptic curve at an $L$ bit security level. It involves selecting an $L$ bit prime $p$, a Galois field $F_p$, a curve $E:y^2 \equiv x^3 + ax + b \mod p$ ($a, b, x, y \in F_p$), and a group $G$ of points $(x, y)$ on $E$, with a generator $P$ of order $q$ with a bit length of $2L$. The TA generates its main secret key $sk_{TA} \in Z_q^*$ and computes the corresponding public key $pk_{TA} = sk_{TA} \cdot P$. The TA then assigns a long-term identity $ID_{cloud}=r_{cloud}^1.P \mod q \in G$ to the server, where $r_{cloud}^1 \in Z_q^*$. The TA also creates five hash functions $h_1()-h_5()$ that produce hashed scalar digests $\in Z_q^*$ from inputs of $\{0, 1\}$ of any length. Two pseudo-point identities $\in G$, $SID_{cloud}^1$ and
\begin{equation}\label{eq3}
SID_{cloud}^2=ID_{cloud} \oplus h_1(sk_{TA} \cdot SID_{cloud}^1, pk_{TA})~mod~q
\end{equation}
are also generated for the server. From these pseudo identities $SID_{cloud}=\langle SID_{cloud}^1|SID_{cloud}^2 \rangle$, alias scalar keys
\begin{equation}\label{eq4}
sid_{cloud}=r_{cloud}^2 + h_2(SID_{cloud}) \cdot sk_{TA} \mod q \in Z_q^*
\end{equation}
are created. The TA can track the server's identity via:
\begin{equation}\label{eq5}
ID_{cloud}^{track}=h_1^{-1}(sk_{TA} \cdot SID_{cloud}^1, pk_{TA}) \cdot ID_{cloud}~mod~q.
\end{equation}

When a vehicle registers with the TA, it receives all elliptic curve parameters except the TA's main secret key $sk_{TA}$. The vehicle $V_i$ is also assigned a long-term identity $ID_{V_i}$, pseudo identities $SID_{V_i}^1$ and $SID_{V_i}^2$, and alias keys $sid_{V_i}$, computed similarly to $ID_{cloud}$, $SID_{cloud}^1$, \eqref{eq3}, and \eqref{eq4}, with different random numbers $r_{i}^1$ and $r_{i}^2 \in Z_q^*$. The TA keeps track of the generated identities for accountability, using $ID_{V_i}^{track}$, calculated as in \eqref{eq5}. The $Twin_{V_i}$ is stored on the server as $\langle ID_{V_i}, SID_{V_i}=\{SID_{V_i}^1|SID_{V_i}^2\}, sid_{V_i}, ID_{V_i}^{track}, f_{V_i}, v_{V_i} \rangle$.

The vehicles perform a safety task of size $S_{V_i}$ bytes. To secure the task at a timestamp $t_{V_i}$, the vehicle signs the hashed task. $V_i$ dynamically generates secret and public keys $sk_{V_i} \in Z_q^*$ and $pk_{V_i}=sk_{V_i} \cdot P \in G$ to ensure anonymity. Using the first-tier alias identities $SID_{V_i}^1$ and $SID_{V_i}^2$, $V_i$ extracts second-tier dynamic alias identities via $h_3()$ and $h_4()$ as:
\begin{equation} \label{eq6}
h_{3,4}(SID_{V_i}|t_{V_i})=h_3(SID_{V_i}^1|t_{V_i})|h_4(SID_{V_i}^2|t_v^i).
\end{equation}
Hashing increases the randomness of the generated signature $\sigma_{V_i}$. Using the alias key $sid_{V_i}$, dynamically generated keys, and hashed alias identities:
\begin{equation} \label{eq7}
\sigma_{V_i}=sid_{V_i}+h_5(pk_{V_i}|h_{3,4}(SID_{V_i}|t_v^i)|S_{V_i}|t_{V_i}).sk_{V_i}~mod~q
\end{equation}
represents $V_i$'s scalar signature on task $S_{V_i}$. Thus, the secured signed task is defined as $d_{V_i}=\langle S_{V_i}, t_{V_i}, pk_{V_i}, SID_{V_i}, \sigma_{V_i} \rangle$.

The IBC-authenticated task $d_{V_i}$ requires signing $c_{{sign}_{V_i}}$ cycles/byte and verification $c_{{verify}_{V_i}}$ cycles/byte computation resources. If the task is processed locally at the vehicle $V_i$, the execution latency $T_{V_i}$ is computed as:
\begin{equation}\label{eq9}
T_{V_i}={d_{V_i} \cdot (c_{{sign}_{V_i}}+c_{{verify}_{V_i}})}/{f_{V_i}}.
\end{equation}

However, vehicles typically prefer to offload their IBC-authenticated computation tasks to the cloud server, leveraging its higher computational power, especially within the 6G framework that supports high-speed data transfers. When the task is offloaded, the cloud first verifies the authenticity of $d_{V_i}$. After checking the freshness of the received timestamp $t_{V_i}$, the server ensures that:
\begin{equation} \label{eq10}
\begin{aligned}
\sigma_{V_i} \cdot P \stackrel{?}{=}& SID_{V_i}^1 + h_5(pk_{V_i}|h_{3,4}(SID_{V_i}|t_v^i)|S_{V_i}|t_{V_i}) \cdot pk_{V_i} \\
& + h_2(SID_{V_i}) \cdot pk_{TA}
\end{aligned}
\end{equation}
holds. If so, the computation time for the task in the cloud is determined by the time needed to verify the task as in \eqref{eq10}, using dedicated cloud computing resources $f_{cloud}^{V_i}$. Once verified and processed, the cloud signs and communicates the result to the vehicle $V_i$. The total time for this process is:
\begin{equation}\label{eq11}
\begin{aligned}
T_{V_i cloud}=&\frac{d_{V_i}}{T_{icloud}^{upload}} + \frac{d_{V_i} \cdot (c_{{sign}_{V_i}}+c_{{verify}_{V_i}})}{f_{cloud}^{V_i}} + \\
& \frac{d_{V_i}}{T_{i cloud}^{download}} + 2\frac{D_{cloud}}{s},
\end{aligned}
\end{equation}
the calculation includes the task uploading time to the cloud, processing on the server, downloading the result back to the vehicle, and propagation delay, considering the distance to the cloud $D_{cloud}$ km and the speed of light $s$ m/sec.

As illustrated in \eqref{eq9} and \eqref{eq11}, the task computation time can vary depending on whether the task is processed locally on the vehicle or offloaded to the cloud server. Consequently, the total task execution latency for $n$ vehicles is:
\begin{equation}\label{eq12}
\begin{aligned}
T_{total}=\sum_{i=1}^{n}\left[(1-\hat x_{V_i})T_{V_i} + \hat x_{V_i}T_{V_i cloud}\right],
\end{aligned}
\end{equation}
where the decision to compute the task locally or remotely is denoted by $\hat x_{V_i} \in \{0, 1\}$.

\subsection{IBC Authentication: Correctness and Security Proofs}
The IBC authentication scheme can be validated for correctness, as the verification of signed tasks at any receiving end from any sending end is successful when the following conditions are met:
\begin{equation}\label{eq13}
\begin{aligned}
&[SID_{V_i} \notin revoked]\,\cap\,[h_2(SID_{V_i}) = Correct]\,\cap \\
&[h_3(SID_{V_i}^1 | t_{V_i}) = Correct]\,\cap\,[h_4(SID_{V_i}^2 | t_{V_i}) = Correct]\,\cap \\
&[h_5(pk_{V_i} | h_{3,4}(SID_{V_i} | t_{V_i}) | S_{V_i} | t_{V_i}) = Correct]\,\cap \\
&[verify_{(P, pk_{TA})}(\sigma_{V_i}) = valid].
\end{aligned}
\end{equation}

Regarding security, the IBC framework ensures resistance to replay attacks using timestamps $t_{V_i}$. It prevents man-in-the-middle, modification, impersonation, and masquerading attacks by applying hashing functions $h_1()$ to $h_5()$. These features protect against any adversary manipulating or compromising the network's legitimate entities, especially in the 6G vehicular networks, where high-speed and reliable communication is paramount. The scheme also demonstrates existential unforgeability against an Adaptive Chosen Message Attack (CMA), where an attacker attempts to forge a signature. Since IBC signatures are already established as secure primitives \cite{al2018lightweight}, further proof is unnecessary. To rigorously evaluate the performance of this secure IBC-authenticated offloading mechanism, particularly the total offloading latency $T_{total}$ across all vehicles \eqref{eq12}, a DRL model is developed below.
\section{DRL-Based Assessment Model Development}\label{hybrid}
We develop a PPO-DRL agent to minimize the total authenticated offloading latency, $ T_{total} $, as expressed in \eqref{eq12}, by making decisions regarding  offloading and resource allocation, subject to the following constraints:
\begin{align}
&\hat x_{V_i}\in\{0, 1\}\quad\quad1\leq i\leq n\, \text{and}\tag{\ref{eq12}a}\\
&\sum_{i=1}^n \hat x_{V_i}f_{cloud}^{V_i} \leq F_{cloud}\quad\quad 0 < f_{cloud}^{V_i} \leq F_{cloud}. \tag{\ref{eq12}b}
\end{align}

The formulation of the DRL model defines the environment, state, agents, actions, rewards, and policy. Our environment, represented by the twin layer, includes $Twin_{V_i}$ and the upload/download data rates, $ T_{icloud} $. The state of the environment reflects the current conditions. The state vector $ State_{cloud} $ captures these features for each vehicle:
\begin{equation}\label{eq:edgestate}
\begin{aligned}
State_{cloud}=\{d_{V_i}, c_{sign_{V_i}}, c_{verify_{V_i}}, f_{V_i}, T^{upload/download}_{icloud}\}.
\end{aligned}
\end{equation}
This state vector is fed into the DRL agent, which includes an actor to propose actions and a critic to evaluate them. The actor operates with both discrete actions $ \hat x_{V_i} $ and continuous actions $ f_{cloud}^{V_i} $, forming our action space:
\begin{equation}\label{eq:edgeaction}
Action_{cloud}=\{\hat x_{V_i}, f_{cloud}^{V_i}\}.
\end{equation}
For $ Action_{cloud} $, the agent receives a reward from the environment, defined as $Reward_{cloud}=-T_{total}$, where the reward is designed to minimize the total latency across all vehicles.

The actor has a neural network that takes $State_{cloud}$ as input and has three output layers; one uses a sigmoid activation for offloading decisions $ \hat x_{V_i} \in \{0, 1\} $ and two generate the mean and standard deviation for resource allocation {\scriptsize$ f_{cloud}^{V_i} $}. The actor's policy $ \pi_\theta $, with parameters $ \theta $, uses the categorical distribution to choose discrete actions and a Gaussian distribution for continuous ones. The log probabilities of these actions are then computed as follows:
\begin{equation}\label{eq:v}
\begin{aligned}
log(p(\pi_\theta))&=log(p(\text{Categorical}(\hat x_{V_i}))) +\\ &log(p(\text{Normal}(\text{mean}(f_{cloud}^{V_i}),\text{std}(f_{cloud}^{V_i}))).
\end{aligned}
\end{equation}
The critic assesses the actor's actions by estimating the state value. The critic neural network processes $ State_{cloud} $ through fully connected layers similar to the actor's architecture. The resulting state value is used to compute the advantage function $\hat{A}$, quantifying the actions' effectiveness. The advantage $ \hat{A} $ is the difference between the critic's estimated state value at time $t$ and the value of the subsequent state:
\begin{equation}\label{eq:k}
\hat{A}_t = Reward^t + \gamma \cdot state_{value}^{t+1} - state_{value}^t,
\end{equation}
where $ \gamma $ is the discount factor balancing immediate and future rewards. We employ the PPO clipping mechanism to optimize the $ \pi_\theta $ policy while ensuring stable learning and constraining policy updates within a predefined range. The PPO clipping loss function $ L^{Clip}(\pi_\theta) $ is formulated as:
\begin{equation}\label{eq:e}
\begin{aligned}
L^{Clip}(\pi_\theta) = \mathbb{E}[min (\hat{r}(\pi_\theta)\cdot\hat{A}, clip(\hat{r}(\pi_\theta), 1-\hat{\epsilon}, 1+\hat{\epsilon})\cdot\hat{A})],
\end{aligned}
\end{equation}
where $ \hat{r}(\pi_\theta) $ is the ratio of new to old policy probabilities, and $ \hat{\epsilon} $ is the clipping hyperparameter controlling the extent of policy updates. To further enhance stability, we incorporate an entropy term into the loss function:
\begin{equation}\label{eq:a}
\begin{aligned}
L^{Clip}(\pi_\theta) = L^{Clip}(\pi_\theta) - H_{coefficient}\cdot H(\pi_\theta),
\end{aligned}
\end{equation}
where entropy $ H(\pi_\theta) = -\sum_{1}^n p(\pi_\theta) \log (p(\pi_\theta)) $ encourages a balance between exploration and exploitation, leading to more effective and stable learning. 

The PPO-DRL process starts by initializing the environment and the agent. The agent selects actions based on the current state, leading to state transitions, rewards, and termination. The policy is updated by adjusting neural network weights to minimize clipped loss using the reward and next state value. The algorithm iterates through training data four times per iteration over 10,000 iterations of 100 episodes and 100 steps each, ensuring stable learning and convergence. The best results were achieved with a learning rate of 0.003, a 0.08 entropy coefficient, and a discount factor of 0.9, which were used to evaluate our IBC-authenticated offloading approach.\label{best}

\section{Simulations and Results}\label{result}
\subsection{Simulation Environment Settings}
The algorithm is implemented in Python 3.11.3 with PyTorch 2.1.2. Simulations run on a MacBook Pro with a 2.8-GHz quad-core Core i7 processor and 16 GB RAM. The cloud server computation capacity $F_{cloud}$ is 20 GHz. The cloud server can dedicate initial computing capacities per vehicle $f_{cloud}^{V_i}$ uniformly distributed between $U[2, 4]$ GHz, while the vehicle's initial capacity is set at 1 GHz. Network size $n$ varies between 10 and 100 to represent different traffic densities. Vehicle speeds $v_{V_i}$ are fixed at 25 m/s, typical for highway settings. The data rate is set between 100 Mbps and 1 Gbps, the lower end of the 6G spectrum, to account for the challenges of maintaining stable connections at higher frequencies. Our IBC scheme \cite{al2018lightweight} signing phase requires five elliptic curve operations: one Scalar Multiplication (SM), one Modular Multiplication (MM), one Modular Addition (MA), and three hashing functions (HS), with a total of $36,000$ cycles per byte on $L=256$-bit elliptic curves, while verification involves three SMs, two Point Additions (PA), and four HSs, totaling $94,000$ cycles per byte. IBC authentication imposes an overhead of 0.04 KB to 0.3 KB on each $ V_i$ task. We use a small 0.05 KB (typical safety task), a medium 30 KB, and a large 300 KB task size.

\subsection{Results and Discussion}\label{results}
Our analysis focuses on offloading latency (msec) and offloading percentage to assess the effect of IBC authentication on task offloading in 6G-VTNs \cite{bilen2022aeronautical,ariman2015software,ak2021forecasting}. All evaluations use the optimal configurations identified in \ref{best}.

\subsubsection{Average Latency (msec)}
\begin{figure}[!htbp]
\centering
\subfloat[]{{\includegraphics[width=.78\columnwidth,trim=2.5cm 2.5cm 1cm 1cm, clip]{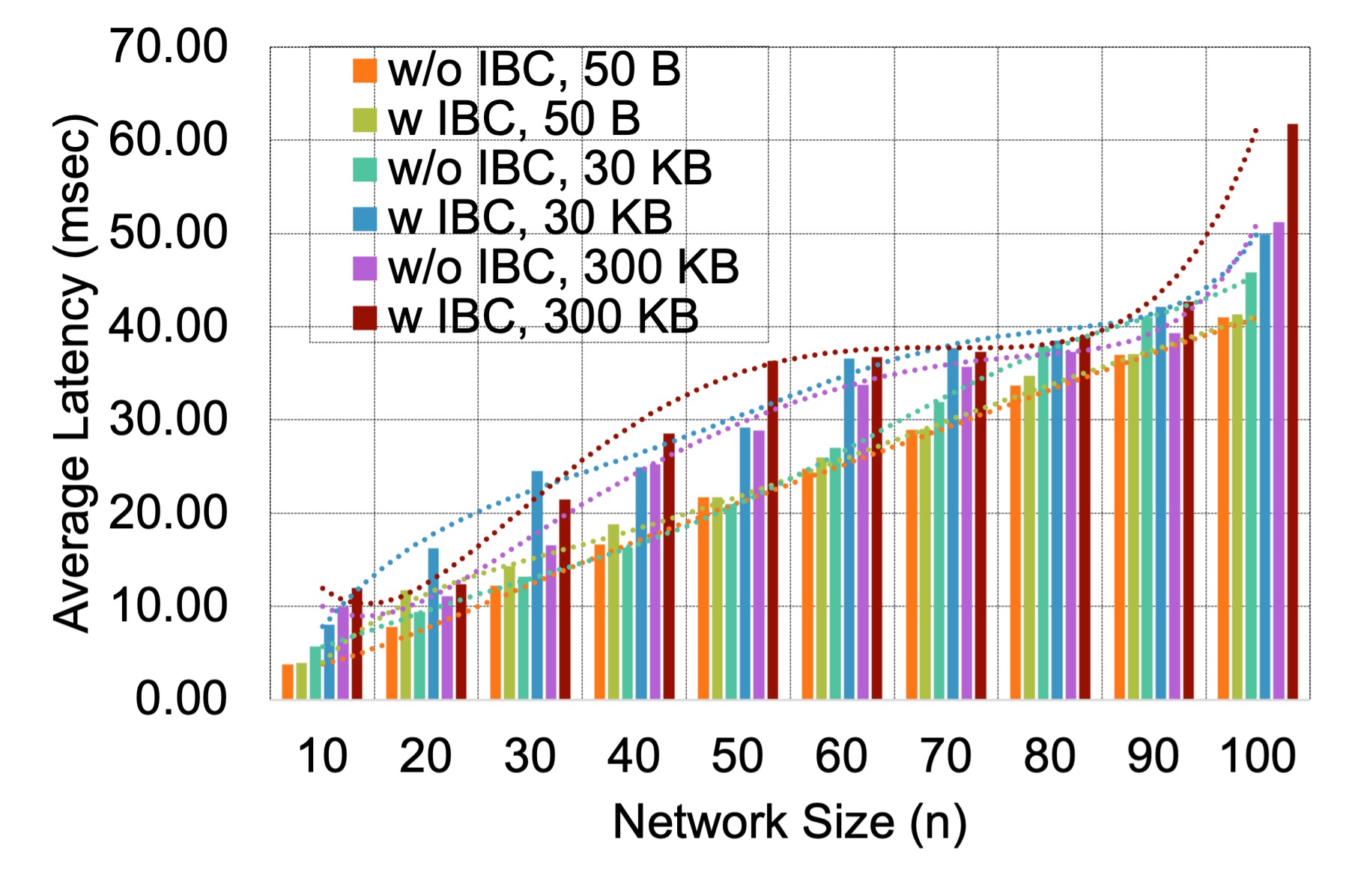}} \label{fig:aa}}\\\vspace{-0.3cm}
\centering \subfloat[]{{\includegraphics[width=.78\columnwidth,trim=2.5cm 2.5cm 1cm 1cm, clip]{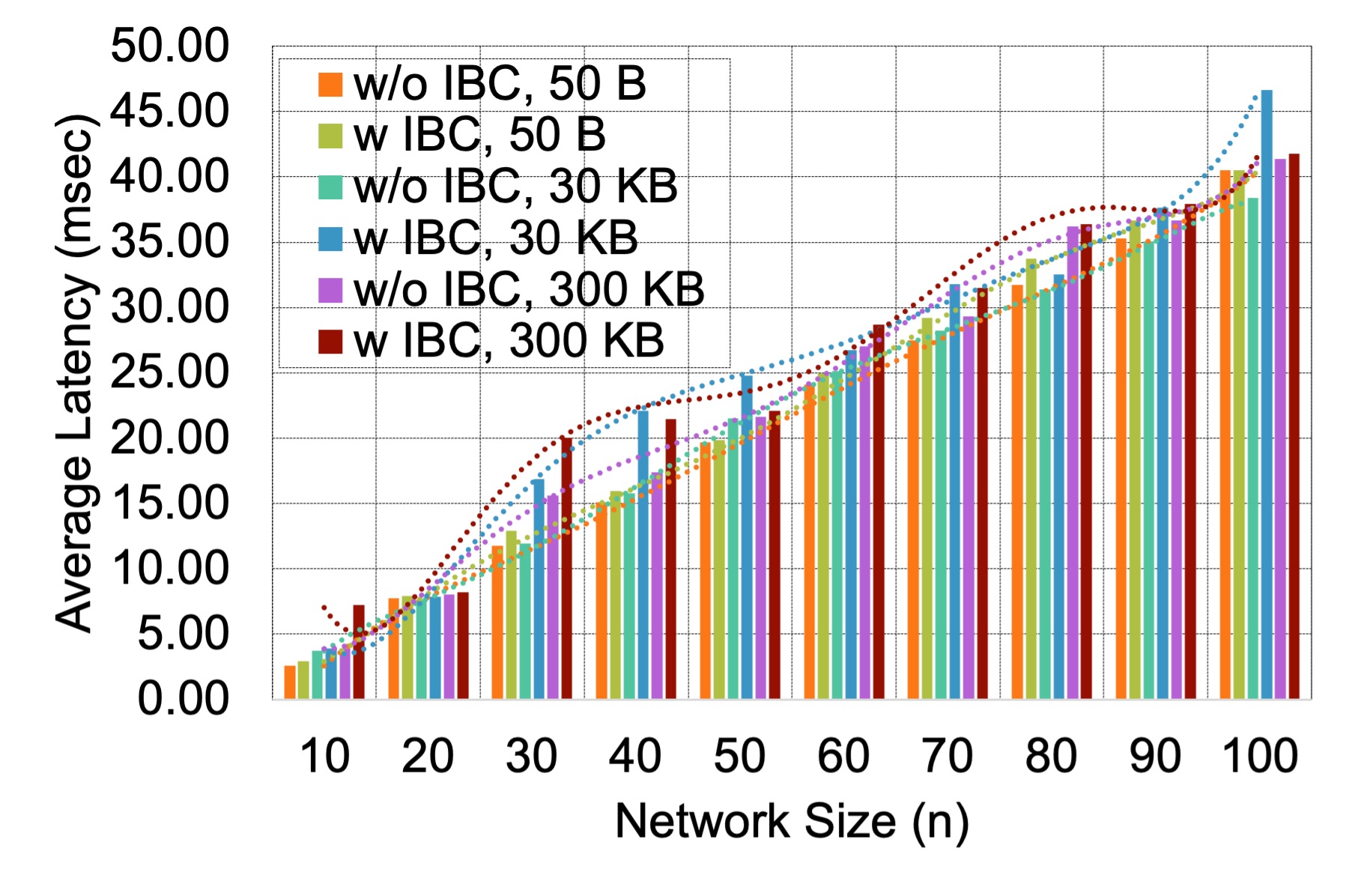}} \label{fig:bb}}\\\vspace{-0.3cm}
\centering \subfloat[]{{\includegraphics[width=.78\columnwidth,trim=2.5cm 2.5cm 1cm 1cm, clip]{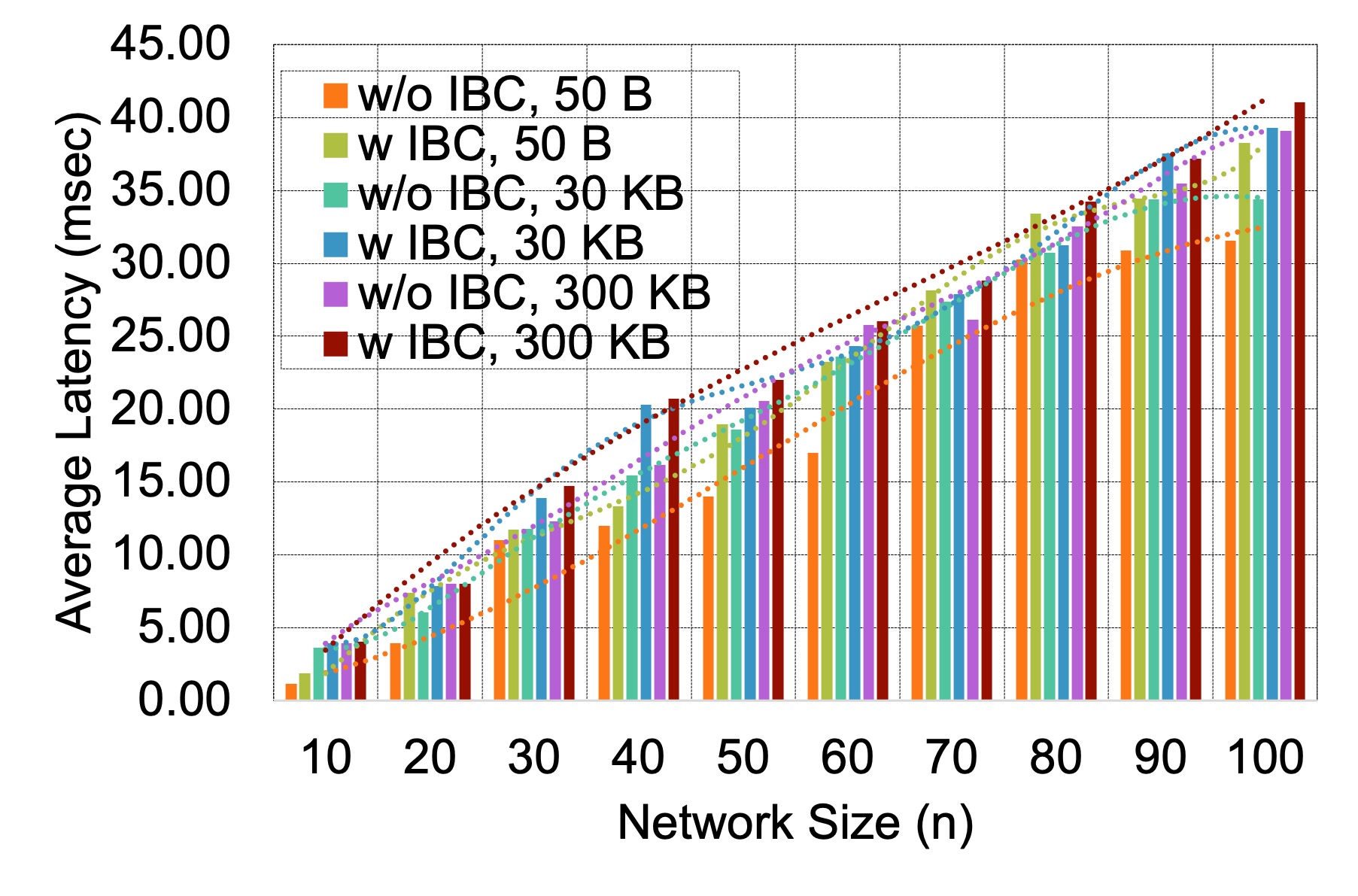}} \label{fig:cc}}
\caption{Average latency (msec) at Different Data Rates. (a) 100 Mbps. (b) 500 Mbps. (c) 1000 Mbps.}\label{fig:abc}
\end{figure}

the average latency is measured against $n$ at three task sizes, with (w) and without (w/o) IBC overhead. In Fig. \ref{fig:abc}\subref{fig:aa}, when the cloud upload/download data rate $T_{icloud}^{up/down}=100$ Mbps, it can be seen that as the network size increases, the offloading latency generally increases. For example, at a small task size of 50 B w/o IBC overhead, the latency increases from 3.76 msec at $n=10$ to 41.00 msec at $n=100$. This effect is due to increased communication overhead and network congestion as more vehicles attempt to offload tasks simultaneously. Similarly, increasing the task size from 50 B to 300 KB results in higher offloading latencies. For $n=10$, the latency w/o IBC increases from 3.76 to 9.99 msec, an increase of more than 165\%. This is expected as larger tasks require more time for both processing and transmission. The introduction of IBC authentication overhead further exacerbates latency, which is particularly noticeable in larger networks and task sizes. For example, at $n=100$ and a task size of 300 KB, the latency w IBC reaches 61.72 msec, compared to 51.23 msec w/o IBC, an increase of 20\%. Increasing the data rate to 500 Mbps in Fig. \ref{fig:abc}\subref{fig:bb} and 1000 Mbps in Fig. \ref{fig:abc}\subref{fig:cc} significantly reduces the offloading latency across all scenarios. For example, with a task size of 300 KB at $n=100$, the latency drops from 61.72 msec at 100 Mbps to 41.75 msec at 500 Mbps and further down to 41.06 msec at 1000 Mbps. This represents a reduction of approximately 30\% when increasing from 100 to 1000 Mbps. These reductions underscore the efficiency gains achieved by higher data rates, which help mitigate the impact of larger tasks and increased network sizes on offloading latency.

\subsubsection{Offloading Percentage}
The offloading percentage represents the proportion of tasks successfully offloaded from the vehicle to the cloud server, with higher percentages indicating more efficient offloading. Table \ref{tab:merged} shows that the percentage of task offload can be initially considered efficient for smaller network sizes $n=10$. For example, with a task size of 50 B, the percentage of offloading with/without IBC starts at 9.73\%. However, introducing IBC significantly impacts this initial efficiency by adding computational overhead, leading to a noticeable reduction in percentage. For example, with IBC, the percentage of offloading for the same 50 B task size at $n=10$ drops from 9.73\% to 4.13\%, indicating that the added security overhead halves offloading efficiency. As network size and task size increase, offloading efficiency decreases further. For instance, for the 50 B task size, at a large network of 100 vehicles, an 89.7\% reduction is seen in the offloading percentage. The larger the network and the larger the task size, the more significant the reduction in offload efficiency. At 300 KB task size, the offloading w/o IBC drops from 6.35\% at $n=10$ to 0.53\% at $n=100$, a reduction of 91. 7\%.

To mitigate the adverse effects of increasing network size and task size on offloading efficiency, we increased the data rate to 500 Mbps and 1000 Mbps in Table \ref{tab:merged}. It is evident that increasing the data rate significantly improves the percentages of task offloading. At 500 Mbps with a task size of 300 KB, the percentage of offloading w/o IBC is 11.85\% at $n=10$, compared to 6.35\% at 100 Mbps, indicating that higher data rates can partially counterbalance increased task size and maintain or improve offloading efficiency. This trend continues at 1000 Mbps, where the percentage of offloading w/o IBC at $n=10$ increases to 17. 18\%, and w IBC, it increases to 15. 65\%. The offload efficiency is further improved for a task size of 300 KB, reaching 12. 85\% w/o IBC and 9.55\% w IBC. Higher data rates give the network greater capacity to handle larger tasks more efficiently, leading to better performance. However, in larger networks, the beneficial effect of higher data rates decreases as task sizes increase due to the compound computational and transmission overhead accompanying larger tasks.

\begin{table*}[!htbp]
\centering 
\caption{Offloading Percentage at 100 Mbps, 500 Mbps, and 1000 Mbps Data Rates}
\resizebox{\textwidth}{!}{\begin{tabular}{ccccccccccccccccccc}\cline{2-19}
&\multicolumn{6}{c}{100 Mbps Data Rate}&\multicolumn{6}{c}{500 Mbps Data Rate}&\multicolumn{6}{c}{1000 Mbps Data Rate}\\\cline{2-19}
&\multicolumn{6}{c}{Task Size (B)}&\multicolumn{6}{c}{Task Size (B)}&\multicolumn{6}{c}{Task Size (B)}\\\cline{2-19}
&\multicolumn{2}{c}{50 B} & \multicolumn{2}{c}{30 KB} & \multicolumn{2}{c}{300 KB} &\multicolumn{2}{c}{50 B} & \multicolumn{2}{c}{30 KB} & \multicolumn{2}{c}{300 KB} & \multicolumn{2}{c}{50 B} & \multicolumn{2}{c}{30 KB} & \multicolumn{2}{c}{300 KB}\\\hline
{n} & {(w/o IBC)} & {(w IBC)} & {(w/o)} & {(w)} & {(w/o)} & {(w)} &{(w/o)} & {(w)} & {(w/o)} & {(w)} & {(w/o)} & {(w)} & {(w/o)} & {(w)} & {(w/o)} & {(w)} & {(w/o)} & {(w)}\\\hline
10  & 9.73  & 4.13  & 8.13  & 3.65 & 6.35  & 3.25& 6.30  & 2.00  & 15.80 & 12.03 & 11.85 & 8.55  & 17.18 & 15.65 & 15.65 & 9.10  & 12.85 & 9.55  \\
40  & 3.58  & 1.73  & 3.72  & 2.97  & 4.22  & 1.57 & 5.18  & 3.56  & 3.56  & 1.65  & 3.35  & 2.88  & 3.16  & 2.67  & 2.67  & 1.74  & 2.60  & 1.63  \\
70  & 2.00  & 0.55  & 2.31  & 1.80  & 1.00  & 0.80 & 2.65  & 1.35  & 2.41  & 1.49  & 1.42  & 1.24  & 2.69  & 1.12  & 1.58  & 1.12  & 2.31  & 0.00  \\
100 & 1.00  & 0.58  & 0.80  & 0.68  & 0.53  & 0.50 & 1.36  & 0.00  & 1.24  & 0.51  & 1.21  & 0.56  & 1.01  & 0.24  & 1.44  & 1.24  & 1.44  & 0.56  \\\hline
\end{tabular}}\label{tab:merged}
\end{table*}

\section{Conclusion and Extensions}\label{conc}
This paper presented a unified framework for task offloading in 6G vehicular networks, integrating lightweight IBC authentication into cloud-based VTNs. Using PPO-DRL, we minimized the offloading process's latency and enhanced resource allocation. Despite IBC's overhead, which reduces the offloading efficiency by up to 50\%, our results showed that increasing the data rate could mitigate these effects and improve the offloading performance by up to 63\%. However, despite such an improvement in large networks and task sizes, the efficiency of offloading is still affected. One future extension will attempt batch verification to lower the verification latency, that is, the total latency. In addition, our goal is to assess more heavyweight authentication, such as group signatures, using the DRL agent developed in 6G vehicular networks.

\section*{Acknowledgment}
This work was funded by the US DOT CARMEN+ UTC Project and the Scientific and Technological Research Council of Turkey (TUBITAK) 1515 Frontier R$\&$D Laboratories Support Program for BTS Advanced AI Hub: BTS Autonomous Networks and Data Innovation Lab Project 5239903.

\bibliographystyle{unsrt}
\bibliography{bibe}
\end{document}